\documentclass[fleqn,usenatbib,useAMS]{mnras}

\usepackage[dvipsnames]{xcolor}
\usepackage{tikz,hyperref}

\definecolor{lime}{HTML}{A6CE39}
\DeclareRobustCommand{\orcidicon}{
	\begin{tikzpicture}
	\draw[lime, fill=lime] (0,0) 
	circle [radius=0.13] 
	node[white] {{\fontfamily{qag}\selectfont \tiny ID}};
	\draw[white, fill=white] (-0.0625,0.095) 
	circle [radius=0.007];
	\end{tikzpicture}
	\hspace{-2mm}
}

\foreach \x in {A, ..., Z}{\expandafter\xdef\csname orcid\x\endcsname{\noexpand\href{https://orcid.org/\csname orcidauthor\x\endcsname}
			{\noexpand\orcidicon}}
}

\usepackage{newtxtext,newtxmath}

\usepackage{booktabs}

\usepackage[T1]{fontenc}
\usepackage{ae,aecompl}

\usepackage{graphicx}	
\usepackage{amsmath}	
\usepackage{booktabs}
\usepackage{times}
\input{epsf}

\title[Reflection in X-ray symbiotic stars]{Reflection physics in X-ray-emitting Symbiotic Stars}

\author[Jes\'{u}s~A.~Toal\'{a}]{Jes\'{u}s~A.~Toal\'{a}\thanks{E-mail:\,j.toala@irya.unam.mx}$^{\orcidA}$\\ 
Universidad Nacional Aut\'{o}noma de M\'{e}xico, Instituto de Radioastronom\'{i}a y Astrof\'{i}sica, Antigua Carretera a P\'{a}tzcuaro 8701, Ex-Hda. San Jos\'{e} de la Huerta, \\
Morelia 58089, Michoac\'{a}n, Mexico
}

\date{\today}

\pubyear{2023}

\begin{document}
\label{firstpage}
\pagerange{\pageref{firstpage}--\pageref{lastpage}}
\maketitle

\begin{abstract}
X-ray-emitting symbiotic stars exhibit a variety of spectral shapes
classified as $\alpha$, $\beta$, $\gamma$, $\delta$ and
$\beta$/$\delta$ types, which have been attributed to different
phenomena such as thermonuclear burning on the surface of the white
dwarf (WD) component, shocks between winds and jets with the red giant
companion's extended atmosphere, the presence of heavily extinguished
hot plasma from the inner region from an accretion disk and/or a
combination of these. However, there is observational evidence that
this classification scheme is not definite and, for example, some
sources change from one type to another within months or years. In
this work, it is proposed that a simple disk-like model can be used to
explain the X-ray properties observed from reflection dominated
symbiotic stars. For this purpose we use the Stellar Kinematics
Including Radiative Transfer ({\sc skirt}) code, which has been
recently upgraded to include radiative transfer from X-ray photons. It
is found that thee properties of the accretion disk (geometry and
density) in combination with the viewing angle can be invoke to
explain the spectral properties of $\beta$, $\delta$ and
$\beta/\delta$ X-ray-emitting symbiotic stars. Spectral variations and
type swaps observed for some X-ray-emitting sources can also be
explained by variations in the disk properties.
\end{abstract}




\section{Introduction}
\label{sec:intro}

X-ray emission from symbiotic stars has been typically used to probe
accretion physics. In such systems, material from a mass-losing red
giant is accreted onto a white dwarf
\citep[WD;][]{Mukai2017}\footnote{A more general definition of a
  symbiotic star has been proposed in the literature by
  \citet[][]{Luna2013}, where other compact objects (neutron stars or
  stellar black holes) accrete material from a red giant companion. In
  this paper we will refer to symbiotic stars as those hosting a WD.}.
It is still not clear what is the exact process in which the accretion
onto the WD is powered \citep[Bondi-Hoyle process, a Roche-lobe
  overflow or a wind Roche-lobe overflow
  channel;][]{Bondi1944,Podsiadlowski2007}, but the formation of an
accretion disk is expected in some sources. In fact, evidence of
accretion disks has been confirmed in the X-ray regime thanks to the
detection of iron lines in the 6.0--7.0~keV energy range
\citep[e.g.,][]{Eze2014}. The 6.7 and 6.97~keV emission lines
correspond to the He- and H-like Fe emission by hot plasma, whilst the
6.4~keV is produced by reflection of hard X-ray photons from the WD or
the accretion disk \citep[see][]{Ishida2009}. Nevertheless, these
emission lines are not detected in all symbiotic stars.

\citet{Murset1997} presented an early scheme to classify symbiotic
stars based on {\it ROSAT} X-ray spectra, which had a relatively soft
spectral range of 0.1--2.4~keV. $\alpha$-type are assigned to
extremely soft sources ($E<$0.4~keV), and are usually attributed to
thermonuclear burning on the surface of the WD \citep{Orio2007},
$\beta$-type symbiotic stars peak at $\sim$0.8~keV and are explained
by the presence of optically thin plasma with temperatures of
$\sim$10$^{6}$ K likely produced by winds and/or jets, but resolved in
only a few sources \citep[see,
  e.g.,][]{Kellogg2001,Kellogg2007,Toala2022}. Finally, the
$\gamma$-type was assigned to sources that exhibited spectra harder
than 1.0~keV, but it is now accepted that these sources might host
neutron stars instead of WD accreting sources \citep[see,
  e.g.,][]{Merc2019}.

Nevertheless, with the advent of the subsequent generation of X-ray
instruments it became clear that symbiotic stars emit hard X-ray
photons, up to 100~keV \citep[][]{Kennea2009}. A fourth category
($\delta$-type) was introduced by \citet{Luna2013} and corresponds to
highly-absorbed, hard X-ray-emitting sources, with a clear presence of
the Fe emission lines in the 6.0--7.0~keV energy range. In addition,
\citet{Luna2013} introduced the $\beta/\delta$-type which was assigned
to sources that exhibit properties of those both types.

\subsection{Reflection models for X-ray-emitting Symbiotic Stars}

While the ($\alpha$, $\beta$, $\gamma$, $\delta$, $\beta/\delta$)
classification scheme is straightforward, it is not definitive. For
example, the recurrent symbiotic star T~CrB evolved from a
$\delta$-type as detected by {\it Suzaku} 2006 observations
\citep{Luna2008} into a $\beta/\delta$-type by 2017 when {\it
  XMM-Newton} re-observed it just after it started a high period of
activity \citep{Luna2018,Zhekov2019}. On the other hand, HM~Sge was
initially classified as a $\beta$-type with a simple spectrum
dominated at 0.8~keV \citep[see fig.~3 in][]{Murset1997}, but the 2016
{\it XMM-Newton} observations detected a new soft component
($E<$0.5~keV) which was attributed to shocked plasma with a
temperature of $\gtrsim3\times10^{5}$~K, very likely produced by a
100~km~s$^{-1}$ jet detected in optical wavelengths
\citep{Corradi1999}. This demonstrates that HM~Sge can not longer be
classified as a $\beta$-type symbiotic star
\citep[see][]{Toala2023}. In addition, it might be also due to
detection limits of the instrument. For example, R~Aqr was classified
as an $\alpha$-type source in the original work of \citet{Murset1997},
but it can be classified as a $\beta$/$\delta$-type after being
observed with the current generation of X-ray satellites \citep[see
  appendix in][]{Toala2023b}. The highly variable nature of symbiotic
stars suggest the classification scheme is misleading, thus
questioning our understanding of the X-ray emission from symbiotic
systems and encourage us to look for a more fundamental scenario to
explain the diversity of X-ray spectra. This situation become more
difficult if one takes into account that only about 60 symbiotic
systems have been detected in X-rays among the almost 300 that have
been confirmed in our Galaxy\footnote{See the statistics in the New
  Online Database of Symbiotic Variables:
  \url{https://astronomy.science.upjs.sk/symbiotics/statistics.html}}.

It is worth mentioning here the similarity between the X-ray spectra
of symbiotic stars and those of active galactic nuclei (AGN), where
reflection physics has been found to be important. Good-quality X-ray
spectra of $\beta/\delta$-type symbiotic systems are very similar to
those of AGN as earlier pointed out by \citet{Wheatley2006}. These
authors compared the {\it ASCA} spectrum of the $\beta/\delta$-type
system CH~Cyg with those of Seyfert 2 galaxies arguing that the X-ray
spectrum of symbiotic systems should be dominated by scattering by an
ionised medium. \citet{Wheatley2006} tired to model the soft component
of the X-ray spectrum of CH Cyg as a result of scattering of the hard
plasma from a photoionised medium around its WD
component. \citet{Wheatley2006} used the {\it absori} model included
in {\sc xspec} \citep{Arnaud1996}, but it does not allow to define a
specific geometry for the absorber \citep[see also][]{Wheatley2003}.

Other authors have attempted models including reflection from
disks. For example, \citet{Ishida2009} used the {\it reflect} model
\citep{Magdziarz1995}, originally tailored for AGN and adopted a thin
distribution of material (a slab) in which reflection from neutral
material is produced. Reflection naturally produces the 6.4~keV Fe
emission line without the need to include an independent (and
arbitrary) Gaussian component in the spectral fit.

Building over the multi-instrument and multi-epoch X-ray data of
CH~Cyg, \citet{Toala2023b} corroborated that indeed a reflection
component is needed to unambiguously fit the 2.0-4.0 keV energy range
of the {\it XMM-Newton} spectrum of this source in addition to the
6.4~keV Fe fluorescent line. However, the reflection component is not
enough to fit the softer range (below 2.0~keV) and low-extinction
thermal plasma are needed. These properties confirm that the physics
behind the production of X-ray emission in symbiotic stars is tightly
correlated with the reflection physics.

This paper is aim to peer at the reflection physics produced by the
presence of an accretion disk around the WD component. We want to
assess the impact produced by different parameters on the observed
X-ray spectra of symbiotic stars and their variability. This paper is
organised as follows. In Section~\ref{sec:methods} we describe the
methodology used in this work. Our results are presented in
Section~\ref{sec:results}. The discussion and conclusions are
presented in Sections~\ref{sec:discussion} and \ref{sec:summary},
respectively.

\section{Methods}
\label{sec:methods}

\begin{figure}
\begin{center} 
\includegraphics[angle=0,width=\linewidth]{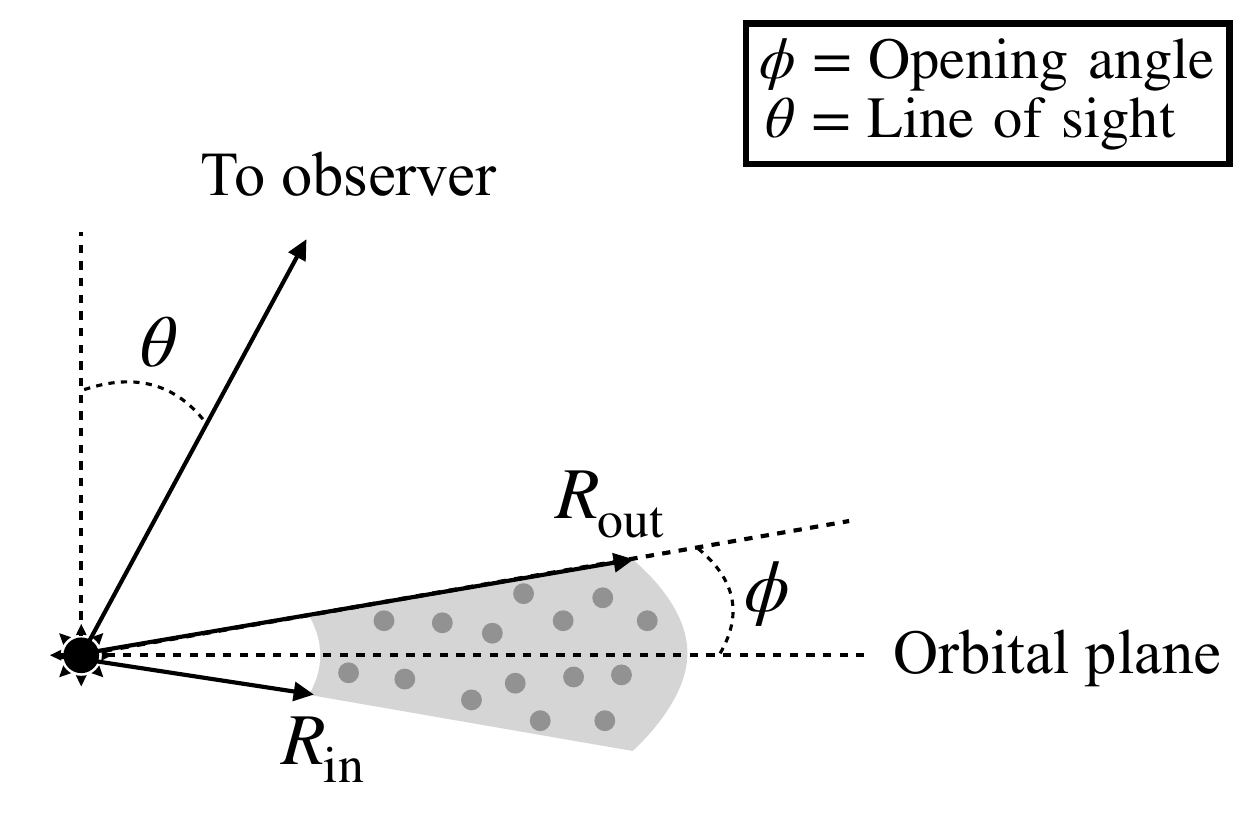}
\caption{Illustration of the geometric scheme (a flared disk) used in our {\sc skirt} simulations. $R_\mathrm{in}$ and $R_\mathrm{out}$ are the inner and outer radii, $\phi$ defines the opening angle and $\theta$ is the line of sight angle towards the observer.}
\label{fig:scheme}
\end{center} 
\end{figure}

For the experiments presented here, we used the Stellar Kinematics
Including Radiative Transfer \citep[{\sc skirt}; version
  9.0;][]{Camps2020} code. {\sc skirt} is a Monte Carlo radiative
transfer code that models the effects of absorption, scattering and
re-emission from a source into a predefined medium. It has been
widely-used to model the CSM around supermassive black holes in AGN
\citep[see, e.g.,][and references
  therein]{GonzalezMartin2023,Trayford2017} and has been recently
extended to include the treatment of X-ray photons by including the
effects of Compton scattering on free electrons, photo-absorption and
fluorescence by cold atomic gas, scattering on bound electrons and
extinction by dust \citep{vMeulen2023}.

Simulations of accreting WDs predict complex toroidal-like density
structures around the binary systems \cite[see for
  example][]{dVBorro2009,Makita2000,Liu2017,Lee2022,Saladino2019}. More
specifically, simulations of the accretion on a WD in a symbiotic
system presented by \citet{Lee2022} predict flared accretion disks
with radius of a few times 0.1~AU. And on the other hand, in the work
of \citet[][]{Liu2017} the density distribution of gas around the
binary system accumulates into toroidal-like structures with external
radii of hundreds of AU, depending on the different mass-accretion
efficiencies and specific angular momentum which are consequences of
the mass ratio of the stars in the binary system.

To illustrate the effects of reflection from the medium around
symbiotic stars, we will use a simple distribution of
material. Reflection calculations using results obtained from detailed
hydrodynamical simulations will be presented in a subsequent
paper. {\sc skirt} allows the user to define a number of built-in
geometries and here we have selected a flared disk characterised by
inner ($R_\mathrm{in}$) and outer ($R_\mathrm{out}$) radii, and
opening angle $\phi$ as illustrated in Fig.~\ref{fig:scheme}. The
flared disk is also characterised by an averaged column density
$N_\mathrm{H}$. A standard value of
$N_\mathrm{H}=5\times10^{23}$~cm$^{-2}$ is used in the calculations
unless specified differently. The models presented here adopt a clumpy
distribution for the flared disk with clump radii of 0.01 AU and a
total number of clumps of 3000 stochastically distributed in the
disk. The temperature of the disk is adopted to be $10^{4}$~K, that
is, completely ionised. The disk clump properties and temperature will
not changed during any of the calculations presented here. All of the
models are run adopting a distance of 1~kpc, consequently, they are
not meant to reproduce any specific object.

\begin{figure}
\begin{center} 
\includegraphics[angle=0,width=\linewidth]{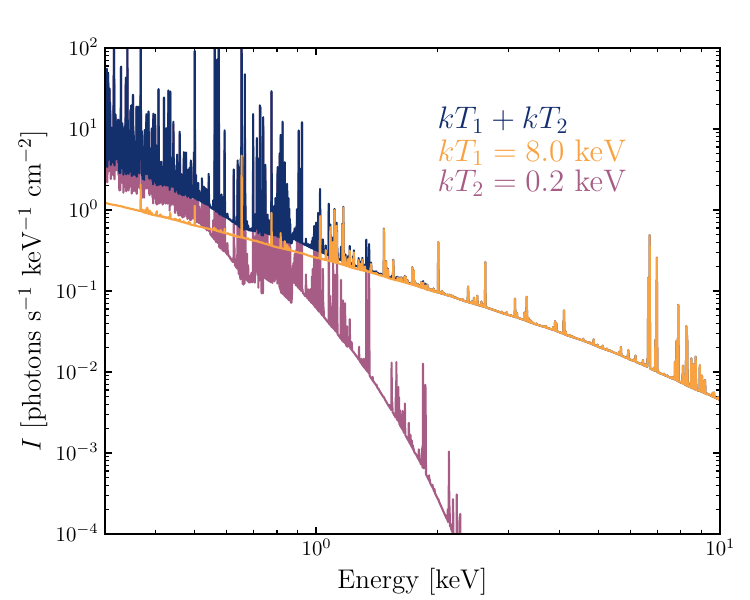}
\caption{Examples of thermal components ($kT_1$=8.0~keV and $kT_2$=0.2~keV) generated with {\sc xspec}. These plasma models are examples of plasma components used as input parameters in the {\sc skirt} simulations. The dark blue line shows the addition of the two $kT_1$=8.0~keV and $kT_2$=0.2~keV  plasma components.}
\label{fig:apec}
\end{center} 
\end{figure}

Another input parameter required in {\sc skirt} is the ionising source
and its properties. In works addressing the AGN properties, the input
source is defined as a power law which is related to the
comptonization process in such systems \citep{Magdziarz1998,Done2012}
and the same approach was adopted by \citet{Ishida2009} for the case
of the dwarf nova SS~Cyg. However, thus far, there is not evidence of
a similar process taking place in symbiotic stars. X-ray studies of
$\delta$- and $\beta/\delta$-type symbiotic stars suggest at the
presence of heavily absorbed, high temperature optically-thin
component \citep{Luna2013}. It is very likely that such emission is
generated at the boundary layer between the inner edge of the
accretion disk and the WD. Matter is transferred from the disk,
loosing its keplerian motion energy, which is deposited on the surface
of the WD. This plasma has been estimated to have temperatures in the
10$^{5}$--10$^{8}$~K range depending on the effectiveness of the
accretion rate \citep{Pringle1979,Patterson1985}, where low plasma
temperatures are associated with high accretion rates and vice versa.

In line with such predictions, the calculations presented here adopt a
shocked plasma model as the input source of X-ray photons, assuming
that the source of X-ray photons is the boundary layer. However, in
our models the source of X-ray photons is adopted to be a point source
at the center of the grid. A plasma temperature of $kT_1$=8~keV
($T_\mathrm{X}\approx9.3\times10^{7}$ K) is adopted in most of the
calculations (unless specified otherwise), more or less consistent
with the hot plasma components reported in other X-ray studies of
symbiotic systems
\citep[e.g.,][]{Ishida2009,Luna2013,Mukai2007,Toala2023b}. An input
table was calculated making use of the {\it apec}
model\footnote{\url{https://heasarc.gsfc.nasa.gov/xanadu/xspec/manual/XSmodelApec.html}}
in {\sc xspec} adopting solar abundances with no extinction (this
effect is consistently calculated by {\sc skirt}). This thermal model
is illustrated in Fig.~\ref{fig:apec} with a yellow line. For
comparison and discussion we also computed other {\it apec} plasma
models with temperatures between 0.1 keV ($\approx10^{6}$~K) and
20~keV ($\approx2.3\times10^{8}$~K). The 0.2~keV emission model is
also plotted alongside the 8.0~keV {\it apec} model in
Fig.~\ref{fig:apec}. All of the calculations were performed adopting
an X-ray luminosity for the {\it apec} models of
$L_\mathrm{X}=2\times10^{33}$~erg~s$^{-1}$ for the 0.3--10.0~keV
energy range.

\begin{figure}
\begin{center}
\includegraphics[angle=0,width=\linewidth]{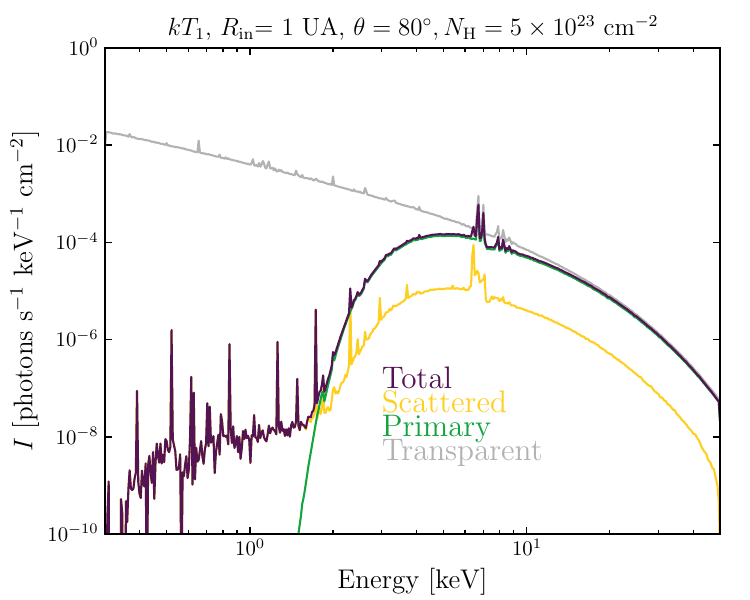}\\
\includegraphics[angle=0,width=\linewidth]{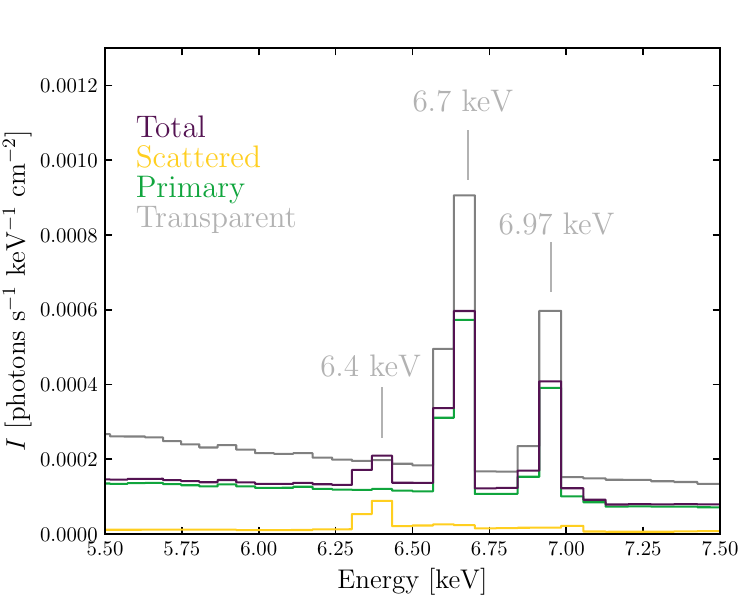}
\caption{Resultant components for a {\sc skirt} simulation with
  standard parameters ($kT_1$=8.0~keV, $R_\mathrm{in}$=1~AU,
  $R_\mathrm{out}$=200~AU, $N_\mathrm{H}=5\times10^{23}$~cm$^{-2}$,
  $\phi=30^\circ$ and $\theta=80^{\circ}$). The bottom panel shows a
  zoom at the three Fe emission lines.}
\label{fig:ct_i80_components}
\end{center} 
\end{figure}

\section{Results}
\label{sec:results}

\subsection{Overall properties}

\begin{figure*}
\includegraphics[angle=0,width=0.5\linewidth]{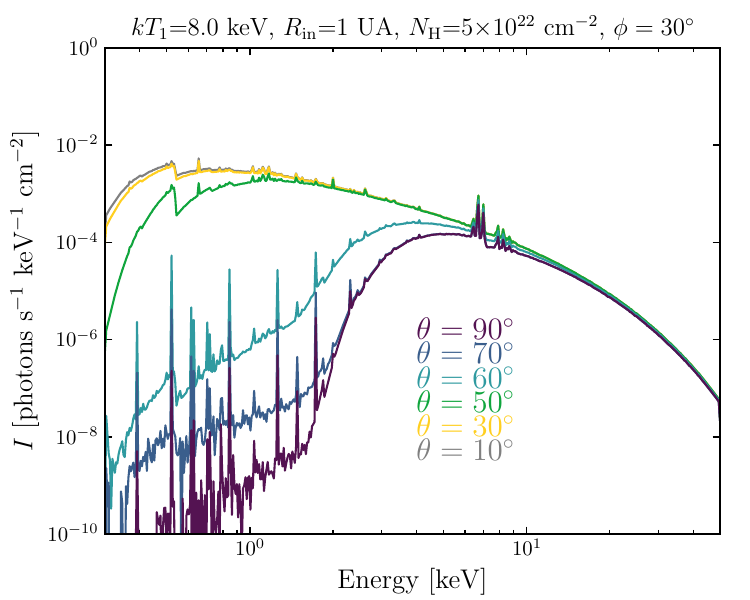}~
\includegraphics[angle=0,width=0.5\linewidth]{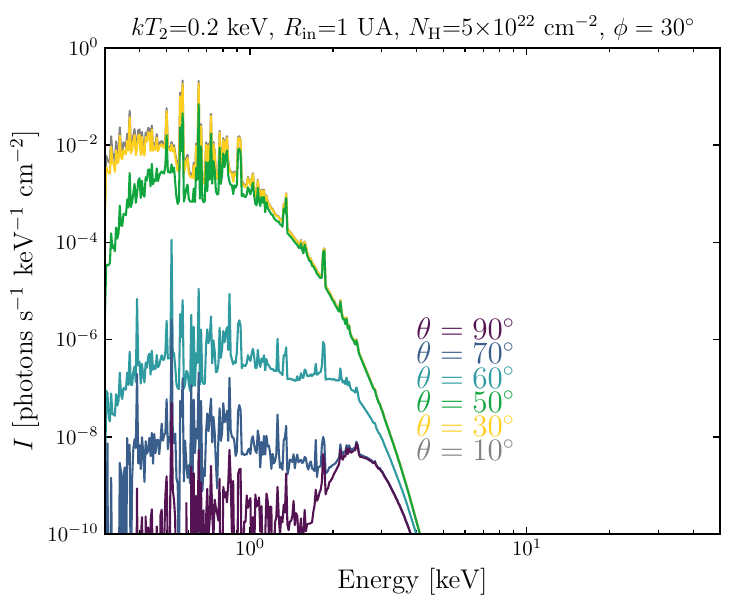}\\
\includegraphics[angle=0,width=0.5\linewidth]{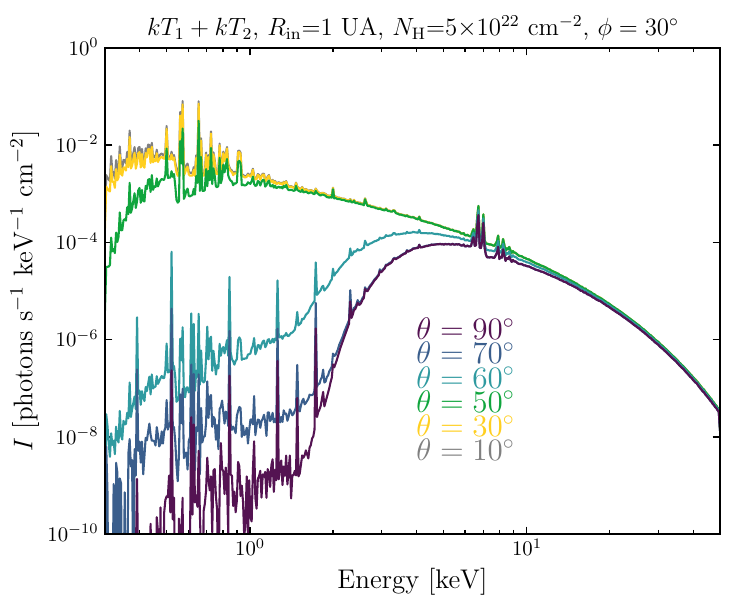}~
\includegraphics[angle=0,width=0.5\linewidth]{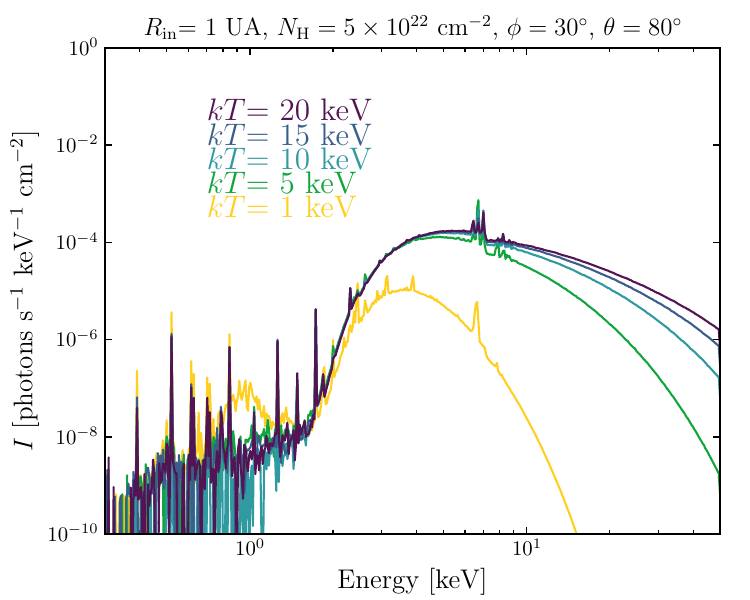}
\caption{Reflection spectra obtained with {\sc skirt} by varying the
  line of sight $\theta$. The top, middle and bottom panels show
  results from adopting input plasma temperatures of $kT_1=8.0$ keV,
  $kT_2=0.2$ keV and $kT_1+kT_2$. In all cases $R_\mathrm{in}$=1~AU,
  $R_\mathrm{out}$=200~AU, $N_\mathrm{H}=5\times10^{23}$~cm$^{-2}$ and
  $\phi=30^\circ$.}
\label{fig:temps}
\end{figure*}

To illustrate the capabilities of {\sc skirt},
Fig.~\ref{fig:ct_i80_components} presents results from a simulation
adopting our standard parameters: $kT_1$=8.0~keV,
$R_\mathrm{in}$=1~AU, $R_\mathrm{out}$=200~AU,
$N_\mathrm{H}=5\times10^{23}$~cm$^{-2}$ and $\phi=30^\circ$. These
results were computed adopting a line of sight of
$\theta$=80$^\circ$. This figure illustrates the different components
reported by {\sc skirt}: the transparent (or unabsorbed) flux, the
absorbed spectrum from the primary source, the scattered spectrum
produced by the disk and the total resultant
spectrum. Fig.~\ref{fig:ct_i80_components} illustrates the
0.3--50.0~keV energy range that can be compared with {\it Chandra},
{\it XMM-Newton} and {\it Suzaku} observations. The bottom panel of
this figure zooms into the 5.5--7.5~keV energy range to show that the
6.4~keV is purely produced by the scattered component, whilst the 6.7
and 6.79~keV Fe emission lines are a consequence of the hot plasma.

It is worth noticing here that a thermal plasma component produces
similar results as adopting a power law distribution as input
model. As a consequence, the emission detected at high energies in all
of our models is thermal in nature. This suggest that accreting,
X-ray-emitting systems are able to produce hard X-ray emission even in
the absence of magnetic fields.

A large number of {\sc skirt} simulations was performed to study the
dependence on the different parameters. Models varying
$R_\mathrm{in}$, $\theta$, $\phi$, $N_\mathrm{H}$ and $kT$ were
performed. We note that varying $R_\mathrm{out}$ did not result in
significantly different synthetic spectra. Thus, this parameter has
been kept fixed in all of our calculations. Further results are
presented in the following subsection.

\subsection{Effects of viewing angle}

We started the {\sc skirt} simulations by evaluating the effect of the
viewing angle. The top-left panel of Fig.~\ref{fig:temps} presents
simulations of the standard model changing the viewing angle $\theta$
from 10$^\circ$ to 90$^\circ$ for models with a single plasma
temperature of $kT_1$=8~keV. This panel shows that as soon as the
inclination angle is large enough ($\theta>50^\circ$ for this set of
models), the disk absorbs most of the soft X-ray emission. As a
consequence, the synthetic spectra resemble those of $\delta$-type
symbiotic stars with negligible contribution from the soft X-ray range
\citep[see for example the case of RT~Cru and
  Hen\,3-461;][]{Luna2007,Luna2013}.

The synthetic spectra of models with large $\theta$ values predict
some emission for energies below 2 keV, but it is so marginal that
real X-ray observations might miss it depending on the exposure time,
the observed count rate and distance to the source. This simple
experiment confirms that the reflection component produced by the disk
plays a major role in shaping the observed spectra of symbiotic stars.

\subsection{Temperature behaviour}

The top-right panel of Fig.~\ref{fig:temps} presents similar
calculations but adopting a plasma temperature of $kT_2=0.2$ keV as
input model. Naturally, these models do not produce significant
emission harder than 3 keV. This panel shows that as soon as the line
of sight is increased, absorption plays an important role in reducing
the total observed flux. Simply because soft photons are more easily
absorbed than hard X-ray emission. High $\theta$ values reduce
considerably the total observed flux. These experiments show that soft
sources (high-accreting systems) will be difficult to be detected for
large $\theta$ values (close to edge-on geometries).

The bottom-left panel of Fig.~\ref{fig:temps} shows calculations
assuming that the central source emits as a two-temperature plasma
emission model, the same hard component of $kT_1=8.0$ keV with an
extra contribution from a soft temperature of $kT_2=0.2$ keV. The
effect of the soft component is appreciated as an excess of emission
lines in the 0.3--1.0~keV energy range, best seen for small $\theta$
values (see bottom panel of Fig.~\ref{fig:temps}), but this is no
longer important for $\theta>50^\circ$. In fact, there are almost no
differences between models in the top-left and bottom-left panels of
Fig.~\ref{fig:temps} for large $\theta$ values. Again, this is purely
due to the fact that the soft X-ray emission is easily absorbed by the
disk material in the line of sight. The top-right and bottom-left
panels of Fig.~\ref{fig:temps} suggest that in those sources that soft
X-ray emission is detected, either we are looking close to pole-on
directions or the soft emission corresponds to extended emission
located outside the disk, where extinction is negligible.

The bottom-right panel of Fig.~\ref{fig:temps} illustrates the impact
of using a harder plasma component. The first thing to notice is that
a plasma model with $kT_1$=1~keV is not able to produce the 6.4~keV Fe
fluorescent line, which is only present for larger input temperature
models. The emission from the 6.97~keV Fe emission line dominates
models with larger $kT_1$ values given that is a higher ionisation
stage. To illustrate the changes in the flux of the Fe emission lines
with temperature we show in Fig.~\ref{fig:iron} a zoom of the
5.5--7.5~keV energy range of this panel (see
Appendix~\ref{sec:iron}). Finally, we note that the increase of the
plasma temperature is reflected in the slope of the continuum for
energies $>10$ keV. Larger input plasma produce larger flux at higher
energies. Higher input plasma temperatures will thus result in high
hardness ratio values.

\subsection{The impact of the disk properties}

\begin{figure}
\includegraphics[angle=0,width=\linewidth]{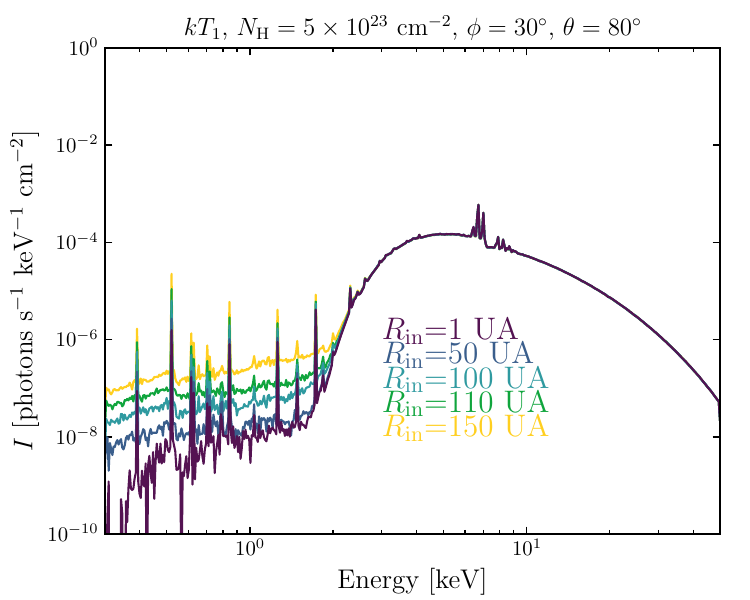}\\
\includegraphics[angle=0,width=\linewidth]{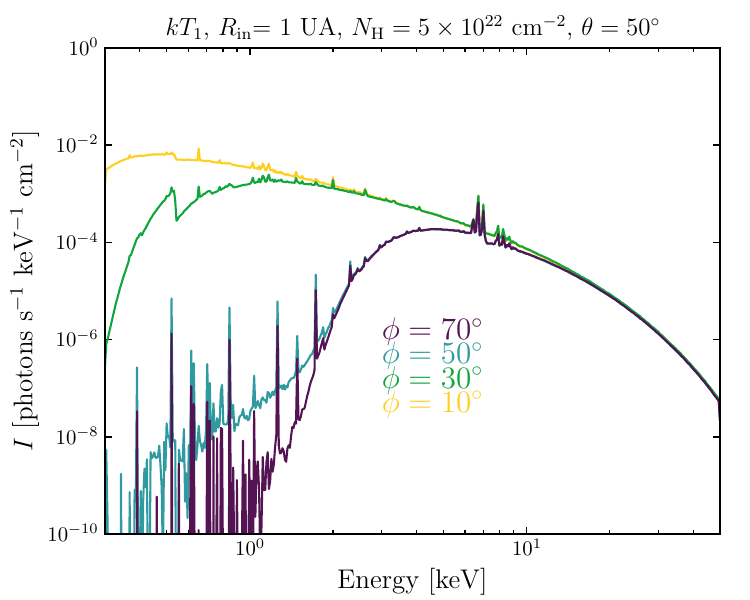}\\
\includegraphics[angle=0,width=\linewidth]{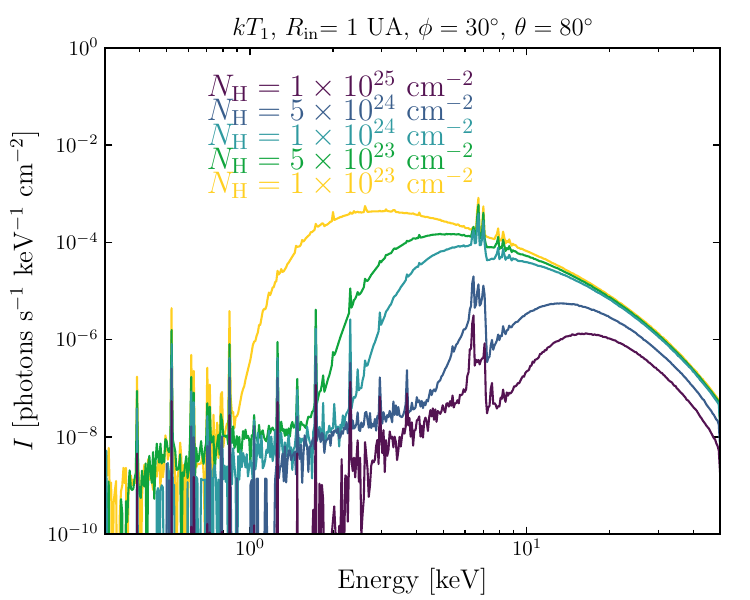}
\caption{Reflection spectra of flared disks obtained from our {\sc
    skirt} simulations varying the inner radius $R_\mathrm{in}$ (top
  left), the opening angle $\phi$ (top right), the column density
  $N_\mathrm{H}$ (bottom left) and the input plasma temperature $kT_1$
  (bottom right).}
\label{fig:others}
\end{figure}

Fig.~\ref{fig:others} presents results of varying other parameters
such as $R_\mathrm{in}$ (top panel), $\phi$ (middle panel) and
$N_\mathrm{H}$ (bottom panel). Varying $R_\mathrm{in}$ produced almost
no difference between models for small $\theta$ (not shown here), but
these are only significant for high $\theta$ values. For example,
Fig.~\ref{fig:others} top panel shows that for $\theta=80^{\circ}$ the
resultant soft ($E<2.0$ keV) X-ray emission is increases by increasing
$R_\mathrm{in}$. That is, the disk is narrower and less material
absorbs soft photons.

The middle panel of Fig.~\ref{fig:others} shows that the variation of
the opening angle $\phi$ produces similar effects as varying the
viewing angle. Large opening angles approximate to a spherical shell
structure around the X-ray-emitting source, which are less likely
under the accretion disk scenario.

Finally, the bottom panel of Fig.~\ref{fig:others} shows the effects
of increasing the disk column density $N_\mathrm{H}$. Same as in the
previous cases, the reduction of the soft X-ray emission by absorption
is increasing by higher $N_\mathrm{H}$ values. However, the most
interesting feature of this panel is the change in the flux of the Fe
emission lines. Particularly note the variation of the fluorescent
line in comparison with those produced by the ionised plasma. A model
with $N_\mathrm{H}=1\times10^{25}$~cm$^{-2}$ is able to completely
extinguish the contribution from the 6.7~keV Fe line. The ratio of the
6.7 keV Fe line over that of the 6.4~keV line increases by reducing
$N_\mathrm{H}$. A figure showing a zoom of the 5.5--7.5~keV energy
range illustrating this situation is also presented in
Appendix~\ref{sec:iron}.

\begin{figure}
\begin{center} 
\includegraphics[angle=0,width=\linewidth]{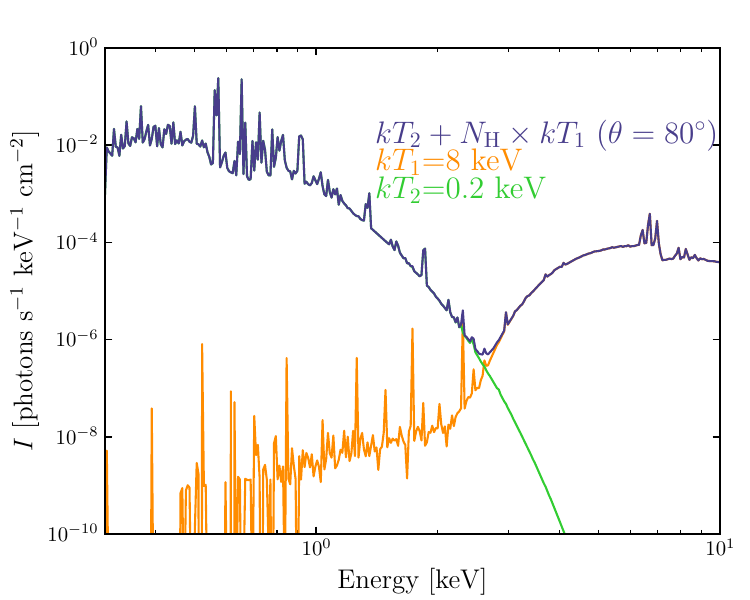}
\includegraphics[angle=0,width=\linewidth]{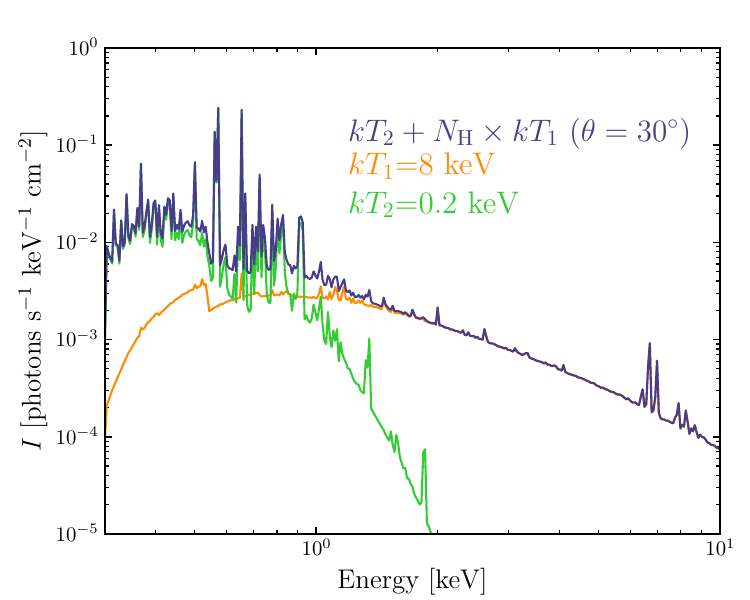}
\caption{Two component {\sc skirt} models. The top panel shows a hard component ($kT_1$=8.0 keV) that has been processed through a flared disk model with $\theta=80^\circ$ as in Fig.~\ref{fig:ct_i80_components} while the soft component ($kT_2$=0.2 keV) has not been extinguished. The bottom panel  presents the same model but with $\theta=30^\circ$.}
\label{fig:2temp}
\end{center} 
\end{figure}

\section{Discussion}
\label{sec:discussion}

It is currently difficult to argue that all symbiotic systems might form an accretion disk, however, in those that is formed, it must definitely have an impact on the X-ray properties as demonstrated here. A relatively simple flared disk model can be used to explain the variety of X-ray properties detected from $\delta$- and $\beta/\delta$-type symbiotic stars, the reflection dominated systems. A disk model including reflection of X-ray photons from a relatively hot source naturally produces the 6.4 keV fluorescent emission line as also discussed by other authors in the past \citep[see, e.g.,][and references therein]{Toala2023b}, but reflection is rarely used when fitting X-ray spectra of symbiotic stars. More importantly, we showed that reflection from a highly-extinguished, shocked hot plasma causes similar effects as that of a power law model (non thermal in origin) for the input source \citep[see][]{Ishida2009}, usually done for AGN.

Close to edge-on viewing angles naturally produce $\delta$-type X-ray-emitting symbiotic stars, very similar to the observed ones \citep[e.g.,][]{Luna2007}. Our simulations show that $\delta$-type symbiotic systems not only require the reflection component seen though a close to edge-on disk-lie structure, but this needs to be produced by relatively hot plasma temperatures ($kT\gtrsim$1 keV). That is, $\delta$-type sources are very likely associated with low-accreting systems.

Including a two-temperature plasma emission model as input does not produce spectra similar to those reported for $\beta$/$\delta$ symbiotic stars where
at least two temperature plasma models are needed: {\it i}) a hot ($kT$=5--20~keV) and heavily extinguished ($N_\mathrm{H} \gtrsim10^{23}$~cm$^{-2}$) component and {\it ii}) the contribution from a soft ($\lesssim$0.2 keV) plasma with a low column density ($N_\mathrm{H} \approx10^{20}$~cm$^{-2}$, very likely due to Galactic column densities). For example, those of CH~Cyg, NQ~Gem, V347~Nor, ZZ~CMi \citep[see][]{Mukai2007,Luna2013,Toala2023}. That is, the soft component of $\beta/\delta$-type sources can not be physically located near the symbiotic system if the spectrum is dominated by reflection as demonstrated in Fig~\ref{fig:temps}. To illustrate the possible situation of $\beta$/$\delta$ sources, Fig.~\ref{fig:2temp} presents calculations of a reflection model with typical values used here and convolved with the contribution from a non-extinguished, soft plasma component with a temperature of 0.2~keV. The figure includes two cases, one with $\theta=80^{\circ}$ (top panel) and another with $\theta=30^\circ$ (bottom panel). Indeed, the model with $\theta=80^{\circ}$ exhibits similar properties as those observed in $\beta/\delta$ sources.

It is thus easily argued that in $\beta$/$\delta$-type objects, the soft X-ray emission must definitely come from an extended source that is no longer extinguished by the disk. That is, outflows (jets and/or winds) produced by the central symbiotic system. The best  example is that of R~Aqr which has been detected to have X-ray-emitting jets \citep{Kellogg2001,Kellogg2007} expanding into hot bubbles. In fact, the jets seem to be feeding the most extended X-ray emission \citep{Toala2022}. Its integrated spectrum including these morphological features, in addition to the central engine, is that of a $\beta$/$\delta$-type symbiotic star as discussed in \citet{Toala2023b}.

The simulations presented here also suggest that high-accreting symbiotic systems that form a disk, that is, sources in which their boundary layer produce soft X-ray photons (some $\alpha$-type sources; see Section~\ref{sec:methods}) can only be observed for small $\theta$ values, that is pole-on directions from the accretion disk. Otherwise, these cannot be detected for edge-on alignments because their soft X-ray emission would be easily extinguished. It is also worth noticing that not all $\alpha$ sources can be directly associated with thermonuclear burning on the surface of the WD. As example, we mention the case of RR Tel where soft X-ray emission can be best explain by shocked-heated plasma very likely due to winds \citep{GonzalezRiestra2013}. Similarly, the soft X-ray component detected in the 2016 {\it XMM-Newton} observations of HM~Sge was produced by shocked-heated plasma from a jet \citep{Toala2023}.

\begin{figure*}
\includegraphics[angle=0,width=0.7\linewidth]{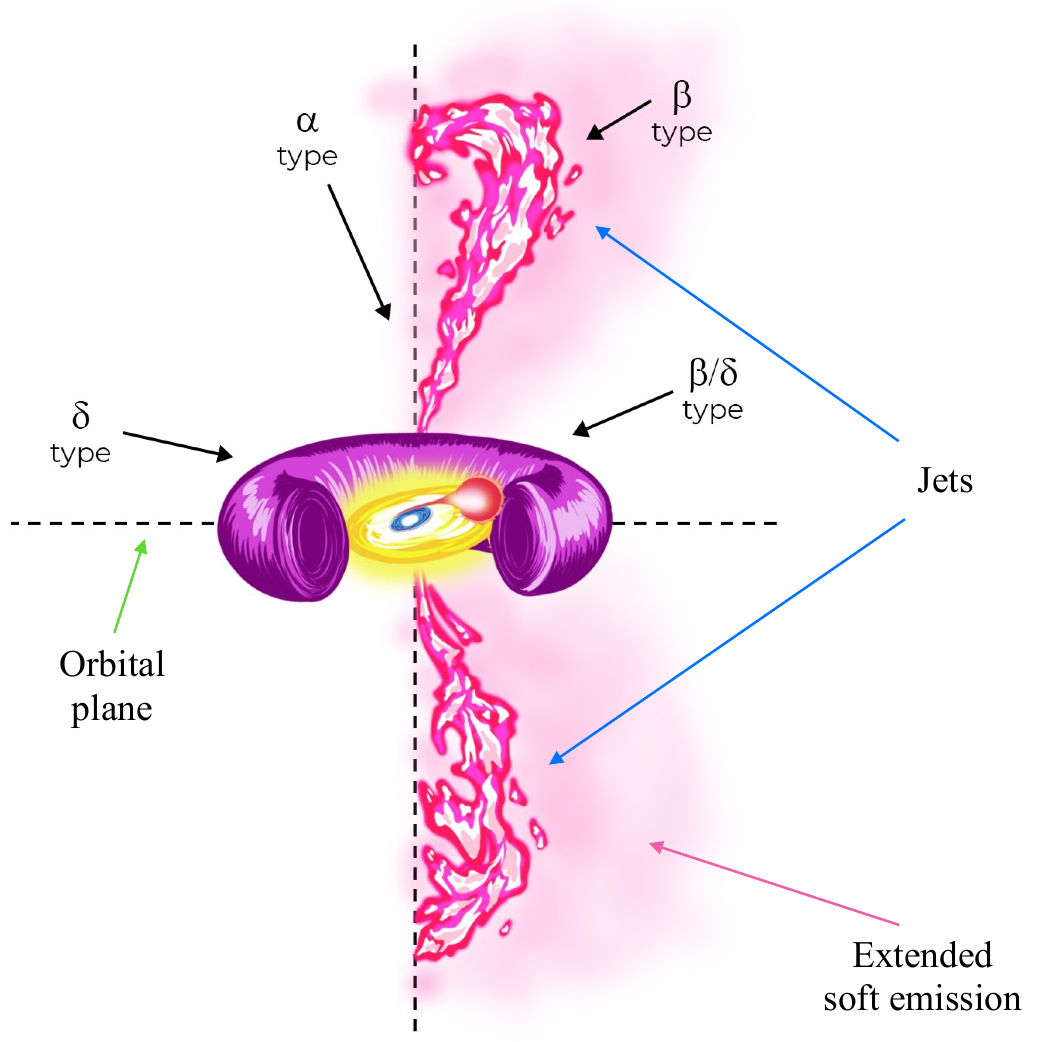}
\caption{Representation of the morphology of the structures around a symbiotic system. The binary and the accretion disk are located at the centre of the sketch. The left side from the vertical (dashed line) represents a system without extended emission, while the right side is that of a system with the presence of a bipolar ejections (jets). The black arrows and labels ($\alpha$, $\beta$, $\delta$ and $\beta/\delta$) represent line of sights. The structures are not to scale.}
\label{fig:sketch}
\end{figure*}

$\beta$-type sources are typically attributed to the presence of shocks produced by  outflows with the extended atmosphere of the red giant companion or the accretion disk. Spectral analysis of $\beta$-type sources do not report high column densities in order to produce acceptable fits \citep[see for example the case of V1329~Cyg;][]{Stute2011}, a feature that suggests that these systems are also observed near pole-on from the orbital plane of the binary system. However, we also note that some sources classified as $\beta$-type symbiotic stars, that is, sources with the peak of emission at about 1~keV also exhibit high energy tails. \citet{Nunez2014} classified Hen~2-87 as a $\beta$ source even after detecting the 6--7~keV Fe feature best seen in the the 2003 {\it XMM-Newton} observations. In addition, their spectral fit required a relatively high $N_\mathrm{H}$ of $\approx10^{22}$~cm$^{-2}$ to obtain a good fit. This points towards the presence of a reflection component that might have been overlooked in $\beta$-type symbiotic systems given the short exposure times of available observations.

There is not much to say regarding $\gamma$-type sources, because they have been defined as symbiotic systems hosting neutron stars \citep[see][]{Merc2019}. However, the evident similarities between X-ray spectra of $\gamma$ sources with those of the $\delta$-type \citep[see for example][]{Bozzo2018,Kaplan2007,Masetti2002,Masetti2007} suggest that similar ideas might be invoked to describe symbiotic systems hosting neutron stars. A situation that is out of the scope of the present paper.

To put these ideas into context, we present in Fig.~\ref{fig:sketch} a sketch of the production of the different X-ray-emitting symbiotic types with viewing angle. To explain the $\beta$ and $\beta/\delta$ sources we require the contribution from extended X-ray-emitting gas such as jets and/or hot bubbles.

There is of course a caveat to the proposed ideas that is worth mentioning, the distribution of material surrounding the symbiotic system. As mentioned before, hydrodynamical simulations have shown that the flared disk is created around the WD component with a relatively small size. Simulations presented in \citet{Lee2022} predict a disk size 0.2~AU, but the rest of the material surrounding the disk and the symbiotic system must have a turbulent toroidal structure extending up to $\sim$100~AU \citep{dVBorro2009,Makita2000,Liu2017,Lee2022,Saladino2019}. Evidently, the next step for this project will be to performed radiative transfer of X-ray photons through density distributions produced by time-dependent hydrodynamic numerical simulations. 
In addition, we rejected large opening angle models because the model geometry is not longer that of a disk. But we note that this parameter also depend on the binary properties as demonstrated by the simulations cited above.

\subsection{Spectral variability}

Multi-epoch observations of symbiotic stars have demonstrated that their spectra change, in some cases dramatically. These clear variations in the spectral shape can also be explained using the results presented in previous sections.

A complex case of spectral variability is that of T~CrB, which went from a $\beta$-type in quiescent phase with a plasma temperature of $\approx$17 keV \citep{Kennea2009} to $\beta$/$\delta$-type with a dominant plasma temperature of $\approx$8 keV after entering a super active state \citep{Luna2018,Zhekov2019}. Different publicly available X-ray spectra obtained from 2017 to 2020 (see Appendix~\ref{app:examples}) show a decrease in soft X-ray emission while the hard X-ray emission increased. \citet{Luna2019} argued that the temperature of the hard component has not changed, only its flux, However, the evident changes in the 2.0--5.0~keV spectral range might suggest that not only the plasma temperature of the hot gas has changes, but other properties as $N_\mathrm{H}$ and very likely $R_\mathrm{in}$. That is, the disk has been experiencing dramatic changes. It is very likely that these have been produce by a disk instability as discussed by \citet{Luna2019,Luna2018}, but we note that such works suggest that the soft X-ray emission is produced very close to the WD surface, but the simulations presented here cannot support this idea because the soft emission would be easily absorbed by the dense accretion disk.
Thus, it is more likely that the soft X-ray emission from T CrB might be produced by winds or jets leaving the densest structure. Another possibility would be that the disk is porous and some of the soft X-ray emission is able to escape before being absorbed. Reflection models tailored to the multi-epoch X-ray observations of T~CrB are most needed in order to confirm such strong assertions (Toal\'{a}, Gonz\'{a}lez-Mart\'{i}n \& Sacchi in prep.).

Another interesting source is the very likely symbiotic system Y~Gem \citep[see, e.g.,][]{Yu2022}. The X-ray data show that reflection is definitely the dominant physics and the 
extreme spectral variations suggest that the disk has experienced dramatic changes in short periods of time (see the bottom panel of Fig.~\ref{fig:examples}). The earliest observation obtained in 2013 did not detect soft X-ray emission ($E<1.0$ keV) and does not exhibit the clear presence of the Fe emission lines. This suggest at a $\beta$-type origin with an extremely large $N_\mathrm{H}$. By 2014 November, the source started exhibiting soft X-ray emission ($E<1.0$ keV) and the appearance of the unresolved contribution from the Fe emission lines. By 2015 the luminosity of the hard component increases whilst maintaining a more or less similar soft flux. Finally, by 2015 the flux of the hard component decreases again. The hard component of Y~Gem do not have the same lower energy range, suggesting that the accretion disk density is changing in very short time scales. A reanalysis of the spectra presented in \citet{Yu2022} is most desirable in order to measure the contribution from the reflection component.

Thus far, the best example of the variation of the three Fe emission lines in the 6.0--7.0 keV energy range is that of CH Cyg. Multi-epoch X-ray observations obtained with {\it ASCA}, {\it Suzaku}, {\it Chandra} and {\it XMM-Newton} obtained from 1994 to 2018 revealed the dramatic flux and equivalent width (EW) changes of the reflection-dominated 6.4 keV line with respect to the 6.7 and 6.79 keV Fe emission lines \citep{Mukai2007,Toala2023b}. Given that the variations in the EW of the 6.4 keV emission line are expected to be the result of the covering angle $\Omega$ and the effective column density $N_\mathrm{H}$ \citep[see, e.g.,][]{Inoue1985}, \citet{Toala2023b} argued that these might be caused by a non-steady mass-loss from the red giant. However, here we also demonstrated that extreme variations in the Fe emission lines can be also produced by varying the plasma temperature of the boundary layer (see Fig.~\ref{fig:iron}). A detailed analysis of the multi-epoch X-ray observations of CH Cyg (and other $\beta/\delta$ sources) is thus needed to unveil the true accretion process and its relationship with the mass-loss ejections from the companion.

\section{Summary and conclusions}
\label{sec:summary}

We presented radiative transfer simulations performed with {\sc skirt} of X-ray photons through a flared disk to study the properties of reflection in symbiotic systems. We assume that the source of X-ray photons is the X-ray-emitting plasma at the boundary layer, the region between the accretion disk and the surface of the WD, which is assumed to be thermal in origin. Models presented here adopted a flared disk geometry that is characterised by inner and outer radii $R_\mathrm{in}$ and $R_\mathrm{out}$, effective column density $N_\mathrm{H}$ and opening angle $\phi$. Simulations were produced by changing the input plasma model $kT$ and were convolved with a viewing angle $\theta$ where 0$^{\circ}$ is pole-on view and $90^\circ$ is edge-on view.

By varying all of the disk parameters, the simulations presented here reproduce a variety of observed properties in the X-ray spectra of symbiotic systems, in particular for those dominated by reflection. We proposed that in such systems the properties of the accretion disk are fundamental to understand the resultant X-ray spectrum in combination with the viewing angle. 

Our main findings can be summarised as follow:
\begin{enumerate}
    \item We assumed that the boundary layer is the source of X-ray photons and that it emits as an optically-thin shocked plasma (i.e., thermal in nature). The interaction of these X-ray photons with the accretion disk produces X-ray spectra of the $\delta$-type similarly to previous analysis of reflection physics adopting a power law (non thermal) as input model. Thus, a magnetic field is not necessary to explain hard X-ray emission from symbiotic systems.

    \item $\delta$-type symbiotic systems are naturally produced by the presence of an accretion disk with close to edge-on viewing angles. The simulations presented here suggest that $\delta$ systems are most likely associated with low-accreting symbiotic stars with high-temperature plasma ($kT>1$ keV) in the boundary layer in order to be detected through their dense disks. Conversely, high-accreting systems, those with soft plasma temperatures in their boundary layers will be difficult to be detected through edge-on viewing angles. 

    \item Using two-temperature plasma models (a hard plus a soft component) as input parameters for the boundary layer cannot reproduce $\beta/\delta$ symbiotic systems. In such cases, the soft X-ray component is easily absorbed by the presence of the disk. Thus, the soft component of the $\beta/\delta$-type sources can only be explain if this emission corresponds to extended components such as jets, winds and/or hot bubbles.

    \item Some soft sources of the $\alpha$ type can be detected if their X-ray emission corresponds to extended X-ray-emitting features such as jets, winds and/or hot bubbles outside the line of sight of the accretion disk. 
    \item Dramatic spectral variations such as the flux and EW of the Fe emission lines in the 6.0--7.0 keV energy range can be  explained by the variable structure of the accretion disk. The intensities of the 6.4, 6.7 and 6.79 keV Fe emission lines are tightly correlated to the increase in the plasma temperature of the boundary layer and to the density of the accretion disk. 
    
    \item Our models predict that the slope of the very hard X-ray emission ($E>10$ keV) increases with the plasma temperature of the boundary layer. Higher plasma temperatures are a consequence of low-accreting systems.
    
\end{enumerate}

We note that we have not tailored our {\sc skirt} models to any specific symbiotic system, but the results presented here have helped us unveil the effects of each parameter of the accretion disk and their specific role impacting the X-ray spectra of symbiotic systems. Spectral fits to a sample of X-ray-emitting symbiotic stars is being perform using {\sc skirt} reflection models through neural networks (Toal\'{a} et al. in prep.). Particular effort will be placed in those with multi-epoch observations as they will help peer into the variable nature of the accretion disks.

Future radiative-transfer calculations performed on more realistic models obtained from time-dependent hydrodynamical numerical simulations will be the next step to corroborate the results presented here.

\section*{Acknowledgements} 

The author thanks comments and suggestions from an anonymous referee that improved the presentation of the results and their discussion. The author also thanks O. Gonz\'{a}lez-Mart\'{i}n, R.~Montez Jr. and M.~A.~Guerrero for fruitful discussions that resulted in the ideas presented in this work. Special thanks to B. vander Meulen for teaching the author how to use the {\sc skirt} code and to J. C. \'{A}lvarez from X.i.k. studio for providing the cartoon used in this paper. This work has made a large use of NASA’s Astrophysics Data System (ADS).

\section*{DATA AVAILABILITY}

The data underlying this article will be shared on reasonable request to the corresponding author.








\appendix

\section{Iron lines}
\label{sec:iron}

\begin{figure}
\begin{center} 
\includegraphics[angle=0,width=\linewidth]{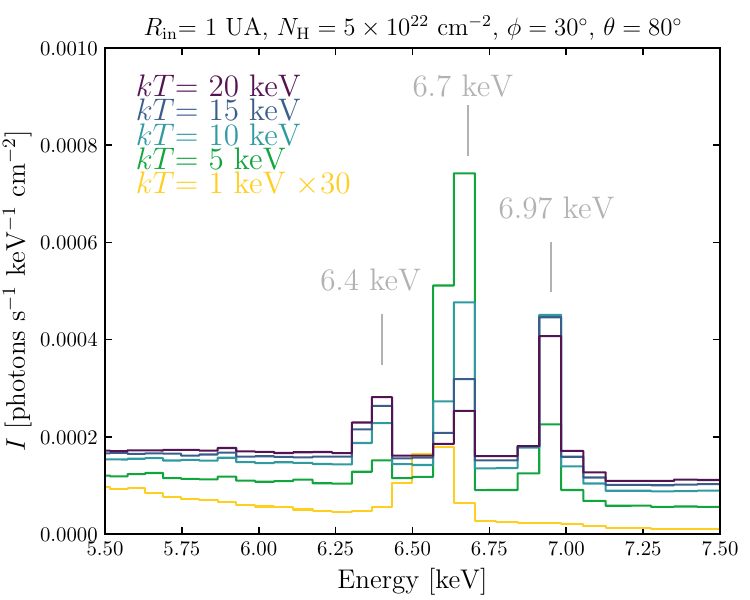}\\
\includegraphics[angle=0,width=\linewidth]{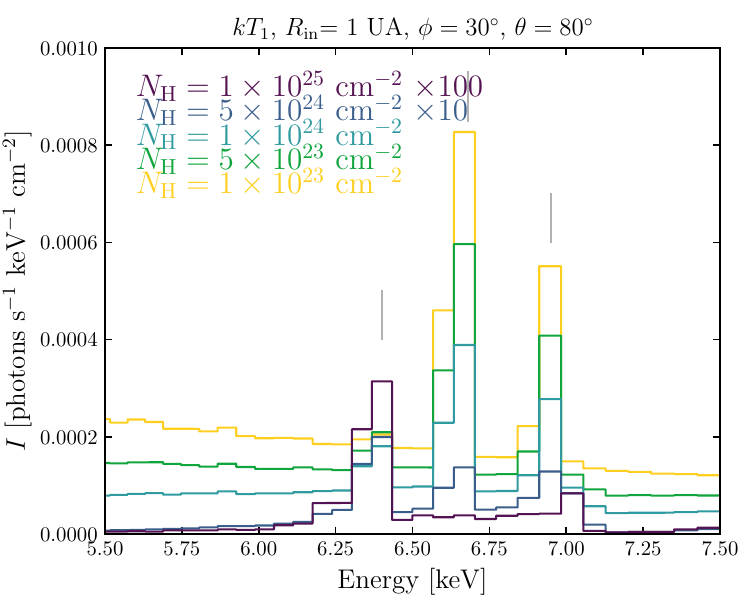}
\caption{Impact on the the three Fe emission lines produced by the plasma temperature $kT$ (top panel) and the column density $N_\mathrm{H}$. The 6.4, 6.7 and 6.79 keV Fe emission lines are marked with grey vertical segments.}
\label{fig:iron}
\end{center} 
\end{figure}

Here we present details of spectra presented in the bottom right panel of Fig.~\ref{fig:temps} and the bottom panel of Fig.~\ref{fig:others} to highlight the changes in the flux of the 6.4, 6.7 and 6.97 keV Fe emission lines. These variations are produced by the plasma temperature $kT$ (top panel of Fig.~\ref{fig:iron}) and the flared disk column density $N_\mathrm{H}$ (bottom panel of Fig.~\ref{fig:iron}).

\section{X-ray spectra of symbiotic stars}
\label{app:examples}

\begin{figure}
\begin{center} 
\includegraphics[angle=0,width=\linewidth]{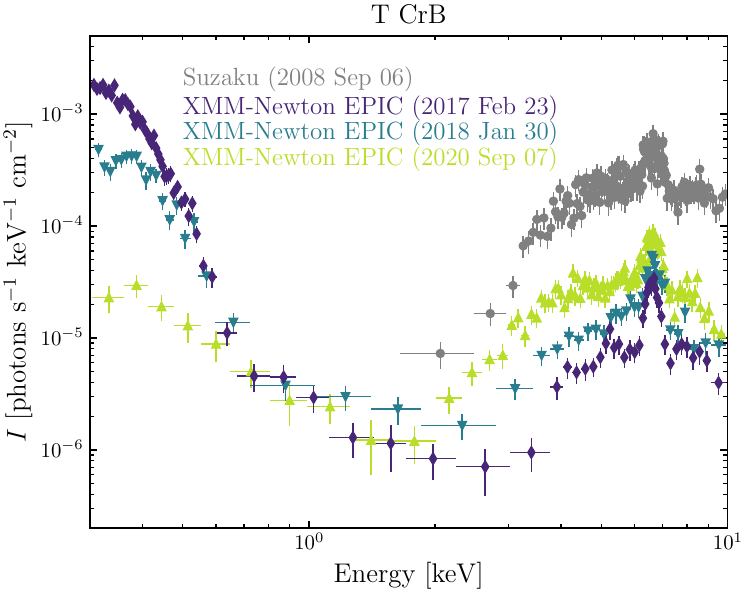}\\
\includegraphics[angle=0,width=\linewidth]{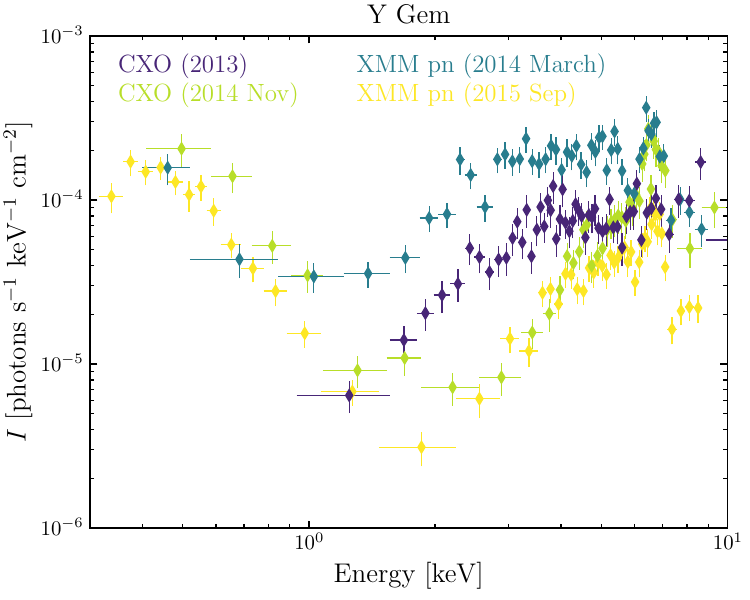}
\caption{X-ray spectra of symbiotic systems obtained from publicly available archives. The panels show spectra from T~CrB (top) and Y~Gem (bottom).}
\label{fig:examples}
\end{center} 
\end{figure}

For means of discussion, here we present publicly available X-ray
spectra of four symbiotic stars. Fig.~\ref{fig:examples} presents
multi-epoch {\it Chandra}, {\it XMM-Newton} and {\it Suzaku}
observations of the $\beta/\delta$-type object T~CrB. We note that the
observations of T~CrB have been analysed in \citet{Kennea2009},
\citet{Luna2018} and \citet{Zhekov2019} except that obtained in 2020.
The bottom right panel shows multi-epoch observations of Y~Gem, a
source that has been studied in the framework of AGB stars with
UV/X-ray excess \citep[see][and references therein]{Ortiz2021},
however, it is very likely that this is a misclassified symbiotic star
as recently suggested by \citet{Yu2022}.



\end{document}